# Electron transmission through a periodically driven graphene magnetic barrier

by

R. Biswas[1], S. Maiti[2] and C. Sinha[1,3]

[1]Department of Physics, P. K. College, Contai, Purba Medinipur, West Bengal- 721401, India

[2]Ajodhya Hills G.S.A.T High School, Ajodhya, Purulia, West Bengal- 723152, India

[3]Department of Theoretical Physics, Indian Association for the Cultivation of Science, Jadavpur-700032, India

Abstract: The kinetic transport of electrons through graphene magnetic barriers is studied theoretically in presence of an external time harmonic scalar potential. The transmission coefficients are calculated in the framework of the non-perturbative Floquet theory using transfer matrix method. The time dependent scalar potential is found to suppress the usual Fabry-Perot oscillations occurring in the transmission through a constant vector potential barrier (corresponding to two oppositely directed δ-function magnetic barriers). Two types of asymmetric Fano resonances (FR) are noted and are discussed for the narrow barrier structure. One of them arises due to the oscillatory mode while the other due to the evanescent mode of the electron wave inside the barrier. In contrast, the oscillating field favours the transmission for rectangular magnetic barrier structure and also exhibits the FR due to the presence of bound state inside the barrier. The characteristic Fano line shape can be tuned by varying the amplitude of the oscillating potential. The detection of such FRs' offers an efficient tool for the identification of the quasi-bound and evanescent extended states inside the barrier not reported in the literature so far, for the case of graphene magnetic barrier structures.
Key Words: Graphene, Magnetic Barriers, Oscillating Field

1. <u>Introduction</u>:- Studies on the magneto-transport properties of a periodically driven graphene[1-3] microstructures is one of the most interesting research topic in recent years, both theoretically and experimentally mainly because of their potential importance for the purpose of magnetic information storage devices as well as for the preparation of magnetic quantum dots to be used in the field of quantum computation. In particular, the current interest in the study of Giant Magneto Resistance (GMR) effect[4,5] modulated by magnetic barriers (e.g., magnetic vector potential) is one of the principal importances in this regards. The use of magnetic barriers[6-8] (instead of the electrostatic one[9,10]) in graphene is one of the efficient pathways to circumvent the severe limitation imposed on graphene (due to the Klein tunnelling effect[10,11]) for its fruitful exploitation in the field of digital electronics.

The kinetic transport in graphene under the action of an external uniform magnetic field exhibits an extraordinary property of unconventional half integer quantum Hall effect[12,13] that could be explained by the existence of relativistic Landau Levels formed by the charge carriers[14]. On the other hand, the use of an inhomogeneous magnetic field introduces the concept of graphene magnetic barrier[6] that totally reflects an incoming electron (thereby suppressing the Klein effect) with energy less than a threshold value related to the total magnetic flux through the barrier[6,8]. Now, when such graphene magnetic barrier is made to be driven by an external oscillating scalar or vector field, many exotic properties could stem from the nano scale multiple field coupling[15-23] and is therefore quite worth studying particularly in the perspective of its high demand in designing electric field (A.C.) tunable graphene based digital nano-devices.

Further, such studies also bear fundamental importance since the underlying quantum physics deals with the interaction between the discrete and the continuum via the exchange of photons. Motivated by all these, the present study addresses the effect of the simultaneous interaction of the chiral Dirac fermions with an oscillating scalar potential and a static magnetic vector potential on the transmission profiles of the electron through graphene based tunnelling microstructures.

In a recent work[24], we studied the electron transmission through a single vector potential barrier (equivalent to two spatially separated δ function magnetic barriers in opposite direction) in graphene based nanostructure in presence of an external laser field. This work is supposed to be the first attempt to study externally driven graphene magnetic barrier structure. However, the δ-function[24-27] type barriers to some extent arise in idealistic situations and are mainly adopted for mathematical simplicity while the realistic situation demands the use of finite width magnetic barriers. In contrast to Ref.[24], the present work deals with a finite width magnetic barrier in presence of an oscillating electric field (instead of laser for the sake of mathematical simplicity). In fact, the problem of laser assisted transmission of the chiral Dirac fermion through a finite magnetic barrier is a bit difficult to solve theoretically, as the final solution becomes almost a formidable task. Apart from the finite width magnetic barrier, we have also studied the case for a pair of δ-function magnetic barriers that exhibit interesting Fabry Perot interference also discussed in the context of laser assisted tunnelling[24]. Here special attention will also be given on the Fabry Perot resonances[28, 29], a widely discussed topic in optics, where a photon bounces back and forth between two coplanar semi transparent mirrors and the successive partially transmitted beams interfere coherently to produce the famous Fabry Perot interference pattern. The present study also aims at studying the effect of an oscillating potential (electrostatic) on the magnetic field induced quasi-bound states[6, 8] in the framework of Floquet non-perturbative approach[30-32], supposed to be the most well known efficient technique for dealing with driven quantum systems.

The importance of the present work regarding the application of a time periodic potential with particular emphasis on photon assisted transport in graphene[33-38] was already emphasised[24]. In the case of a periodically driven quantum systems, the inelastic scattering channels open up due to the exchange of photons between the tunnelling electron and the oscillating potential. Fano type resonances[39-43] are likely to appear in this context due to transitions between the Floquet sideband states and the bound states both for the δ-function as well as finite width graphene magnetic barriers.

Another aspect is to study the extended states inside a constant vector potential barrier in graphene that are responsible to produce a different kind of the Fano resonances, not yet reported in the literature. The control or manipulation of these quantum states and the transmission profiles in graphene based tunnelling structures are inevitable not only from the theoretical point of view but also for their successful exploitations in device fabrications, e.g., in sensing and switching applications[44].

## 2. Theoretical Model:

The two band Dirac-Weyl Hamiltonian of a monolayer graphene in an external magnetic field described by a space dependent vector potential $\vec{A}(x)$ can be written as,

$$H_0 = v_F \vec{\sigma}.[\vec{p} - e\vec{A}(x)] \qquad (1)$$

where $v_F$ is the Fermi velocity = $c/300$, '$c$' being the velocity of light; $\vec{\sigma} = (\sigma_x, \sigma_y)$ are the Pauli matrices representing pseudo-spin analogous to the original spin; $\vec{p} = -i\hbar(\partial_x, \partial_y)$ is the momentum operator in the graphene plane ($x$, $y$); '$e$' being the electronic charge; the vector potential $\vec{A}(x)$ is uniform along the y-direction but varies along the $x$- direction. The magnetic field is chosen along the z direction, perpendicular to the graphene plane.

For a rectangular magnetic barrier of length $L$, the vector potential profile (Fig. 1(a)) polarized along the y-direction ($\hat{y}$) and the corresponding magnetic field profile (Fig. 1(b)) along the x-direction are respectively given by;

$\vec{A}(x) = -(LB/2)\hat{y}$     for  $x < -L/2$

$$= xB\hat{y} \quad \text{(in units of } B_0 l_0\text{)} \quad \text{for } -L/2 < x < L/2$$

$$= (LB/2)\hat{y} \quad \text{for } x > L/2 \quad \quad 2(a)$$

and $\vec{B}(x) = B\,\hat{z}$ (in units of $B_0$) for $|x| < L/2$

$$= 0 \quad \text{elsewhere}. \quad \quad 2(b)$$

Here $l_0 = \sqrt{\frac{\hbar}{eB_0}}$ is the length scale with a typical magnetic field strength $B_0$; The above potential profile can be created by depositing a ferromagnetic strip on top of the graphene layer[6, 8].

Let us now assume that a time varying sinusoidal scalar potential $V(t) = V_0 Cos(\omega t)$ is applied in the region $|x| < L/2$ (Fig. 1(c)), where $V_0$ and $\omega$ are the amplitude and frequency of the oscillating potential respectively. Thus, in the region II ($-\frac{L}{2} < x < \frac{L}{2}$, Fig. 1(c)), the Dirac fermion satisfies the corresponding time dependent wave equation $H(t)\phi(x,y,t) = i\hbar \frac{\partial \phi(x,y,t)}{\partial t}$, where the time dependent Hamiltonian is given by $H(t) = v_F \vec{\sigma} \cdot [\vec{p} - e\vec{A}(x)] + eV(t)$. Since the Hamiltonian is periodic in time, one can consider the wave function as $\phi(x,y,t) = e^{-iE_F t/\hbar} \Psi(x,y,t)$, $E_F$ being the Floquet quasi-energy. $\Psi(x,y,t)$ is a periodic function in time satisfying $\Psi(x,y,t+T) = \Psi(x,y,t)$, $T$ being the time period of the oscillating field. Assuming the same form of time dependence of the electron in both the sub-lattices, the full wave function in region II may be considered as $\phi(x,y,t) = (\Psi_a(x,y), \Psi_b(x,y))^T f(t) exp(-iE_F t/\hbar)$. Further, the continuity of the Hamiltonian along the y-direction leads to the y-component of wave function as $\sim e^{ik_y y}$. Using all these, the wave equation for the Dirac fermion ends up with two coupled differential equations:

$$\frac{-i}{\Psi_a(x)}\left[\frac{d}{dx} + (k_y + xB)\right]\Psi_b(x) = (E_F - V_0 cos\omega t) + \frac{i}{f(t)}\frac{df(t)}{dt} \quad \quad 3(a)$$

$$\frac{-i}{\Psi_b(x)}\left[\frac{d}{dx} - (k_y + xB)\right]\Psi_a(x) = (E_F - V_0 cos\omega t) + \frac{i}{f(t)}\frac{df(t)}{dt} \quad \quad 3(b).$$

In eqns (3), all the dynamical variables are written in dimensionless form, e.g., $x \to x\, l_0$, $B \to BB_0$, $E_F \to E_F E_0$, , $V_0 \to V_0 V^/$ and $\omega \to \omega \omega_0$, where $\omega_0 = \frac{v_F}{l_0}$, $E_0 = \hbar \omega_0$ and $V^/ = \frac{E_0}{e}$.

Now using the method of separation of variables, one can write

$$(E_F - V_0 \cos\omega t) + \frac{i}{f(t)}\frac{df(t)}{dt} = E \qquad \qquad 3(c),$$

$E$ being a constant.

The coupled equations 3(a) and 3(b) reduce to

$$-i\left[\frac{d}{dx} + (k_y + xB)\right]\Psi_b(x) = E\Psi_a(x) \qquad \qquad 4(a)$$

$$-i\left[\frac{d}{dx} - (k_y + xB)\right]\Psi_a(x) = E\Psi_b(x) \qquad \qquad 4(b).$$

The solution for the time dependent part in Eq.3(c) may be written as $f(t) \sim e^{-i(E-E_F)t - i\alpha \sin\omega t}$ with $= V_0/\omega$. To find the constant $E$, we apply the Jacobi-Anger identity $e^{-i\alpha \sin\omega t} = \sum_{n=-\infty}^{+\infty} J_n(\alpha) e^{-in\omega t}$ so that the periodicity of $f(t)$ leads to the relation $E = E_F - n\omega$. Here $J_n(\alpha)$ is the Bessel function of order 'n', 'n' being the index of photon number (absorbed /emitted). The form of $E$ indicates that the spinor wave functions in eqs. 4(a) and 4(b) satisfy an infinite set of coupled differential equations

$$\frac{d\Psi_b^m(x)}{dx} + (k_y + xB)\Psi_b^m(x) - i(E_F + m\omega)\Psi_a^m(x) = 0 \qquad \qquad 5(a)$$

$$\frac{d\Psi_a^m(x)}{dx} - (k_y + xB)\Psi_a^m(x) - i(E_F + m\omega)\Psi_b^n(x) = 0 \qquad \qquad 5(b)$$

The index 'm' now corresponds to the Floquet side bands arising from the exchange of photon due to the interaction of the Dirac fermion with the external oscillating field. For a particular value of 'm', the above coupled differential Eqns. 5(a) and 5(b) is well known to be equivalent to a pair of decoupled Schrödinger like equations, e.g., for $\Psi_a^m(x)$ one can write

$$\frac{d^2 \Psi_a^m(z)}{dz^2} - \left[z^2 + 1 - \left\{\frac{(E_F + m\omega)^2}{B}\right\}\right]\Psi_a^m(z) = 0 \qquad \qquad (6)$$

with $= \sqrt{B}(x + k_y/B)$. The solution of Eqn.(6) may be given by the Hermite polynomial[8] with the energy spectrum $E_l^m = m\omega \pm \sqrt{2B(l+1)}$ ; 'l' being any positive integer, called the Landau level index and 'm' is an integer that runs -∞ to +∞, called Floquet sideband index. Thus $E_l^m$ corresponds to the energy of the Landau-Floquet level that arises due to the simultaneous application of magnetic field and oscillating electric field on graphene monolayer.

It would be more suitable for the tunnelling problem[6] to express the solutions for $\Psi_a^m(x)$ and $\Psi_b^m(x)$ in terms of the Parabolic cylinder function as

$$\Psi_a^m(x) = A_m\, D_{\nu_m}(x\prime) + B_m\, D_{\nu_m}(-x\prime) \quad \text{and}$$

$$\Psi_b^m(x) = \left(i\sqrt{2B}/(E_F + m\omega)\right)\left[A_m\, D_{\nu_m+1}(x\prime) - B_m\, D_{\nu_m+1}(-x\prime)\right] \quad (7)$$

$D_{\nu_m}(x\prime)$ being the Parabolic Cylinder function[45] with $\nu_m = \frac{(E_F+m\omega)^2}{2B} - 1$ and $x\prime = \sqrt{2B}\left(x + k_y/B\right)$. $A_m$ and $B_m$ are the constant coefficients corresponding to $m$-th sideband.

Now, for the case of a constant vector potential (Fig. 1(d)) in the region II (producing a pair of two oppositely directed and specially separated) δ-function magnetic barriers[24-27], the corresponding solutions may be given by

$$\Psi_a^m(x) = A_m\, e^{iq_m x} + B_m\, e^{-iq_m x} \quad \text{and}$$

$$\Psi_b^m(x) = \left[A_m \frac{q_x^m + i\{k_y+B\}}{E_F+m\omega} e^{iq_m x} + B_m \frac{-q_x^m + i\{k_y+B\}}{E_F+m\omega} e^{-iq_m x}\right] \quad (8)$$

with $(q_x^m)^2 = (E_F + m\omega)^2 - (k_y + B)^2$.

The solutions in regions I and III are of the same form as in our previous work[38].

Finally, matching the pseudospin components at the two interfaces ($x = L/2$ and $x = -L/2$), one can find the transmission coefficient ($T_m$) for the $m^{th}$ sideband of energy $E_m$ and the amplitude of transmission $D_m$ by using the following relation[18, 38]:

$$T_m = \frac{\cos\theta_m}{\cos\theta_0} \left|\frac{D_m}{F_0}\right|^2 \qquad (9)$$

with $\theta_m = \tan^{-1}\left(\frac{k_y}{k_x^m}\right)$ and $(k_x^m)^2 = (E_m)^2 - (k_y)^2$ ; $F_0$ is the amplitude of the wave incident at an angle $\theta_0$ in the region I.

## 3. Results and discussions:

### 3A: Transmission through a pair of Delta function magnetic barriers:

At first we would like to recapitulate the Fabry Perot interference (FPI) for electron transmission through a pair of static δ-function magnetic barriers[24, 29]. The multiple beam (Fabry Perot) interferometry, usually used in optics, is a manifestation of the wave nature of matter. The FPI effect for electron is totally absent in the case of graphene single electrostatic barrier for normal incidence due to the presence of the Klein transmission. On the other hand, for the magnetic barrier in graphene, FPI is noted for the normal as well as for the glancing incidences[27]. It should be mentioned here, that the FPI is observed in the case of above barrier tunnelling when the solutions of the Dirac Weyl equation are oscillatory, both inside and outside the barrier regions. The FPI maxima ($T_c \sim 1$) are observed when the wave vector inside the barrier satisfies the relation $q_x = n\pi/L$, n is an integer and '$L$' being the width of the barrier. As for example, we display in Figs. 2(a) and 2(b) the transmission coefficients ($T_c$) for a single scalar ($V_z = 2$) and vector barriers ($B = 2$) respectively with '$L$' = 20 and $k_y = 0.5$, showing the Fabry Perot resonance – a signature of interference. For a single vector barrier, the transmission is zero for the energy $E < (k_y + B)$ and as such, the origin of the energy scale is taken to be as $E \gtrsim (k_y + B)$. It may be noted from the above figures that the number of FPI maxima for a given energy interval is greater for the vector barrier than those for the electrostatic one, the condition of occurrence for the latter being given by[9] $q_n = \frac{n\pi}{L}$. Further, in the case of a single scalar barrier of height '$V_z$', the $T_c$ at the minima of the FPI increases with the increase in energy more rapidly than those for the case of the vector barrier. This could probably be explained as follows. In the case of scalar barrier, the pseudo-spinor component $\varphi_1(x, y)$ satisfies the Schroidinger like equation with effective potential $V_{eff} = k_y^2 - 2EV_z + V_z^2$ in contrast to the vector barrier for which $V_{eff} = k_y^2 + 2Bk_y + B^2$. Thus the effective barrier height decreases with the increase in energy for the electrostatic barrier (but not for the vector barrier) and the transmission rapidly approaches to unity. For a single electrostatic barrier, the condition for occurrence of the FPI for electron transmission is given by $|E - V_z| > k_y$. Thus for the scalar barrier, the FPI can be observed under the below barrier condition (Inset of Fig. 2(b)) but not for the vector barrier.

In order to comprehend the effect of the external scalar time periodic potential on the electronic transport through a graphene vector barrier, we display in Fig. 3(a), the $T_c$ for $\omega = 0.5$ and $V_0 = 1$ (other parameters same as Fig. 2(a)). Since the under barrier transmission coefficient is of ~ of $10^{-6}$ even in presence of the oscillating field, we plot only the above barrier transmission throughout the paper. The Fabry Perot (FP) oscillations in the energy dependent transmission are noted to be modified appreciably in presence of the oscillating field as compared to the static condition. Interestingly, although the magnitude of the modification is oscillatory at lower incident energy, it is monotonic for higher energy of incidence on the vector barrier. As regards the qualitative aspect, it is noted that at low incident energy, the modified FP oscillation is in phase with that under the static condition while at higher incident energy, there is a $180^0$ phase difference between the two. In contrast, for the case of oscillating scalar barrier (vide Inset of Fig. 3(a)), the FP oscillations appear as a step like pattern at lower incident energy whereas the oscillation almost dies out approaching $T_c \sim 1$ at the higher value of the incident energy.

Fig. 3(b) reveals that the FP oscillation is very much sensitive to the amplitude of the oscillating field. At lower $V_0$, the FP oscillation is suppressed only at the lower energy end of the transmission spectrum. While, with the increase in $V_0$, the energy range of the modulated oscillation gradually increases. This is quite justified since the low energy electrons have longer time of interaction with the oscillating field than that for the higher energy one. Ultimately with the increase in $V_0$ to a value $\gtrsim 0.4$, the oscillation in the FP transmission almost ceases within a small energy window (node) at the lower energy end (around E = 4.63, for $V_0 = 0.5$ from Fig. 3(b)) of the spectrum. As we go on increasing the amplitude, the number and width of nodal windows increase and they systematically move towards the higher energy end of the spectrum (e.g., for $V_0 = 1.0$ two nodal widows are at E = 4.15 and 5.45).

In order to study the effect of the frequency of the oscillating field $\omega$ on the FP transmission, we plot the total transmission $T_c$ for $\omega$ = 0.1, 0.2, 0.3 and 0.4 respectively in Figs. 4(a-d). It may be mentioned here that due to the presence of the Bessel function whose argument involves the frequency ($\omega$), the overall modification of the FP transmission oscillates with $\omega$. As is noted from the Fig. 4(a), for $\omega = 0.1$, the oscillating field suppresses the FP transmission throughout the range considered here and the oscillation almost dies out above '$E$' ~ 8. With the increase in $\omega$, the frequency of oscillation increases (vide Figs. 4(a) and 4(b)) and a nodal window appears around '$E$' ~ 4.66 for $\omega = 0.3$ (vide Fig. 4(c)). Further, the number of nodes increases with the increase in $\omega$ (e.g., two for $\omega = 0.4$, as displayed in Fig. 4(d)), keeping the mean transmission almost the same as that noted in the absence of the oscillating field.

The modulation of the FP oscillation induced by the oscillating field is also sensitive to the sign of the angle of incidence on the vector barrier (VB). It is already known that the isotropy of transmission through the electrostatic (scalar) barrier disappears due to the application of the vector potential[6]. The basic differences between the FP transmission (for the static vector barrier) for +ve and –ve $k_y$ are as follows: i) for the –ve $k_y$ the separation between the FP maxima increases with increasing energy more rapidly than that for the +ve $k_y$ and ii) the $T_c$ at the FP minima is greater for the latter than that for the former. Now in presence of the oscillating potential (vide Fig. 5 and Inset), the modification of the low energy above barrier FP transmission for the +ve $k_y$ is more oscillatory as compared to the case for the –ve $k_y$. The suppression of transmission is more prominent for the +ve glancing incidence than that for the negative one.

Another salient feature of the present time dependent magneto-tunnelling problem is the appearance of the Fano resonance in the transmission spectra that arises due to the coupling between the continuum and the discrete states via photon exchange. The type of the Fano spectrum was already reported for the graphene based electrostatic well / barrier structures[33, 34]. The appearance of the Fano line shape bears the signature of the discrete quasi-bound states inside the well/barrier of the quantum tunnelling system. The quasi bound states can be obtained under the static condition when the solution is oscillatory inside the well/barrier and evanescent outsides. In case of a single vector barrier structure in graphene, the presence of the quasi-bound state is restricted by the condition $|k_y + B| \leq |E| \leq |k_y|$ while the positions of the quasi-bound states are obtained from the transcendental equation $Tan(q_x L) = k_x q_x / [E^2 - k_y(k_y + B)]$. For the parameters $B = 1$, $k_y = -1$ and $L = 2$, the only quasi bound state occurs at $E_b = 0.5145$. Thus the conservation of energy leads to the position of the Fano resonance to appear at $E = E_b + n\omega$ ( n being a positive integer ), which is quite legitimate as shown in Fig. 6(a). The figure also reveals that the two photon processes (the FR's at higher energy) are less probable than the single photon processes. The effect of changing the amplitude of the oscillating field ($V_0$) can be observed from Fig. 6(b) which indicates that the static barrier exhibits only one extended peak in the energy region considered, unlike the case of the Fabry Perot interference, observed at larger width of the vector barrier. With the increase in $V_0$, the peak to dip ratio decreases systematically for the single photon processes, in sharp contrast to the two photon processes where the aforesaid ratio increases. This indicates that the multi photon processes become more probable with the increase in the amplitude of the field.

Figs. 7(a) and 7(b) plot the total transmission coefficients for $B = 1$, $k_y = -0.5$ and $L = 2$. For these parameters the static vector barrier does not support any quasi bound states with the restrictions (as mentioned above) of oscillatory nature inside the barrier and evanescent outside. In spite of that, we find the appearance of the Fano line shapes in the transmission spectrum, not yet reported in the case of realistic time dependent scalar barrier[33, 34]. The positions of the asymmetric resonances ($E = 0.817$, $0.872$ and $0.929$, in Fig. 7(a)) indicate that there must be some other type of discrete state with energy $E_b \sim 0.2$ in the static vector barrier. The existence of this state contradicts the previously mentioned condition $|k_y + B| \leq |E| \leq |k_y|$. This new type of discrete state probably arises due to the evanescent mode of propagation both inside and outside the vector barrier, unlike the case of the previous one (-ve $k_y$). The condition for occurrence of such extended state is $|E| \leq \min\{|k_y + B|, |k_y|\}$ and the corresponding energy is governed by the equation $exp(-2q_x L) = [(k_x + q_x)^2 - B^2]/[(k_x - q_x)^2 - B^2]$ for $q_x^2 = (k_y + B)^2 - E^2$. The energy conservation relation implies that the energies corresponding to the extended states are $E_b = 0.217$, $0.222$ and $0.229$ respectively for $\omega = 0.6$, $0.65$ and $0.7$ as displayed in Fig. 7(a), i.e., the new kind of state is blue shifted ( arising due to AC Stark shift ) with the increase in frequency $\omega$. So far as the effect of $V_0$ is concerned, it may be noted from Fig. 7(b) that the width of the FR increases with the increase in $V_0$. Although the peak-to-dip ratio for the single photon processes remains almost constant for all $V_0$, it increases more rapidly with $V_0$ for the two photon processes.

Finally, we display in Figs. 8(a-c) the nature of the transmission through different Floquet sidebands arising due to the scattering of the charge carriers from a time dependent graphene vector barrier. Fig. 8(a) displays the result for the transmission coefficients for the single photon absorption ($T_{+1}$), the single photon emission ($T_{-1}$) and for no photon ($T_0$) processes for the system with $L = 20$, $k_y = 0.2$, $B = 2$, $V_0 = 1.0$ and $\omega = 0.6$. The figure indicates that at low incident energy ($E < 5$), the variation of the $T_c$ lies almost within the same limit (positions of the envelope maxima though relatively shifted) while for higher energy ($E > 5$), the probability of transmission via the photon exchange processes gradually dies out. All the electrons transmit through the central band with reduced amplitude of Fabry Perot oscillation. The two photon processes ($T_{+2}$ or $T_{-2}$) also occur at lower energy but with lesser probability as compared to the single photon processes (vide inset of Fig. 8(a)). The situation just reverses with the decrease in $\omega$, as may be noted from Fig. 8(b). Thus at higher frequency and/or lower amplitude of the oscillating potential, the over barrier electrons prefer to transmit through the central band as compared to the higher Floquet side bands. To study the nature of the Fano resonance in the side band transmission, we present in Fig. 8(c) the results for $T_0$ and $T_{\pm 1}$ in the case of narrow ($L = 2$) vector barrier structure. It may be noted that the central band (with no photon exchange) exhibits contrasted FR in comparison to the side bands. On the other hand, the photon absorbed process depicts a less contrasted FR at lower energy while the photon emitted process exhibits the same at higher energy. This is probably due to the fact that the photon emitted transmission starts at a comparably higher energy due to unavailable electronic states inside the barrier.

3B: Transmission through a driven rectangular magnetic barrier-

In order to study the effect of an external time dependent scalar field on the transmission of Dirac Fermion through a rectangular magnetic barrier we have plotted in Fig.

9(a) the angular variation of the total side band transmission ($T_c$) at different values of $V_0$, e.g., 0.0, 0.2, 0.5 and 0.7. It may be mentioned[6] that under the field free condition the electron can transmit through the barrier with finite probability only for the angle of incidence ($\theta_0$) satisfying the condition $Sin\theta_0 = Sin\theta + \frac{BL}{E}$, so that for positive glancing incidence no transmission occurs for $E \leq BL$ (as is noted from Fig. 9(a)). In contrast, for negative glancing incidence, the above momentum conserving relation is satisfied with allowed values of the angle of transmission θ and therefore the $T_c$ is found to be appreciable for $E$ either greater or less than the product of $LB$. Regarding the effect of time varying field Fig. 9(a) reveals that for $\theta_0 < 0$ the $T_c$ increases appreciably with increasing $V_0$ around the maximum of the angular transmission profile keeping the angular half width at half maxima almost constant. Thus the effect of the field is to increase the quality factor of resonance and hence the sharpness of the angular transmission through the magnetic barrier. On the other hand, the $T_c$ becomes finite for +ve glancing incidence under the application of the field. This is particularly because under oscillating field the Floquet side bands with energy $E + n\omega$ play the relevant role in transmission dynamics. The scattering through the side bands arising from photon absorption processes are now allowed by the $k_y$ conservation and the $T_c$ increases with the increase in $V_0$ even for the +ve angle of incidence.

Finally, Fig. 9(b) depicts the energy dependence of the $T_c$ for rectangular magnetic barrier at different frequencies of the oscillating field. It would be worthwhile to mention here that for the present parametric condition the static magnetic barrier supports only one bound state at the energy $E_b = 0$. Therefore the transmission profile under oscillating field should display the Fano resonances at incident energies satisfying the relation $E = E_b + n\omega$. This is verified from the results in Fig. 9(b) where the FR's are noted at the energies E = 0.8, 1.2 and 1.6 for the case of ω = 0.4 corresponding to the emission of two, three and four photons respectively. The FR due to the single photon emission is absent as it would correspond to the evanescent transmission in the outgoing channels. Comparing the results for the other frequencies it may be pointed out that photon exchange processes are more probable at lower frequencies of the oscillating field and the transmission with the exchange of higher number of photons are less pronounced than that for lower number of photons.

## 4. Conclusion:

The salient features of the present work where the magneto – tunnelling properties of the Dirac fermions in a graphene based microstructure are studied under the action of an oscillating time dependent potential could be summarized as follows.

The FPI pattern due to electron tunnelling through a pair of static δ-function magnetic barriers suffers amplitude modulation by the application of the oscillating time dependent potential, particularly for large '$L$'. At comparatively higher frequency and amplitude of the oscillating field the, oscillation completely ceases around certain small energy windows producing nodes in the FP transmission. In contrast for small '$L$' asymmetric Fano resonances are noted instead of the Fabry Perot resonance. Two types of the FR are noted – one arising due to the quasi-bound state for which the electron wave is oscillatory inside the barrier and the other due to the extended state that is evanescent inside the barrier. The latter FR is dynamic in character in the sense that it is blue shifted with the increase in frequency ω. Although the position of the FR is controlled by ω, its shape is sharply dependent on the amplitude $V_0$. Regarding the side band transmissions, the high energy electrons prefer to tunnel through the higher Floquet bands with decreasing frequency ω. On the other hand, for rectangular magnetic barrier, the Landau-Floquet energy levels are derived for the first time. Regarding transmission, it is noted that the time varying field favours transmission both for the +ve and –ve glancing incidence and the energy dependent transmission exhibits the Fano resonances confirming the position of bound states inside the barrier.

References:

[1] K. S. Novoselov et.al., Science 306 (2004) 666.

[2] K. S. Novoselov et.al., Nature 438 (2005) 197.

[3] Y. Zhang, Y. W. Tan, H. L. Stormer and P. Kim, Nature 438 (2005) 201.

[4] X. Yang, et. al. Appl. Phys. Lett. 108 (2016) 252401.

[5] W. T. Lu, Appl. Phys. Lett. 105 (2014) 012103.

[6] A. De Martino, L. Dell' Anna and R. Egger, Phys. Rev. Lett. 98 (2007) 066802.

[7] L. Oroszlany, P. Rakyta, A. Kormanyos, C. J. Lambert and J. Cserti, Phys. Rev. B 77 (2008) 081403(R).

[8] M. R. Masir, P. Vasilopoulos, A. Matulis and F. M. Peeters, Phys. Rev. B 77 (2008) 235443.

[9] M. I. Katsnelson, K. S. Novoselov and A. K. Geim, Nat. Phys. 2 (2006) 620.

[10] C. Bai and X. Zhang, Phys. Rev. B **76** (2007) 075430.

[11] O. Klein, Z. Phys. 53 (1929) 157.

[12] F. Guinea, M. I. Katsnelson and M. A. H. Vozmediano, Phys. Rev. B 77 (2008) 075422.

[13] K. S.Novoselov et. al., . Science **315** (2007)1379.

[14] C. M. Lu et. al. Apl. Phys. Lett. 96 (2010) 212101.

[15] C. H. Park, L. Z. Tan and S. G. Louie, Physica E: 43 (2011) 651.

[16] S. Ghosh and M. Sharma, J. Phys.: Condens. Matter 21, 292204 (2009).

[17] R. Biswas, A. Biswas, N. Hui and C. Sinha, J. Appl. Phys. 108 (2010) 043708.

[18] M. A. Zeb, K. Sabeeh and M. Tahir, Phys. Rev. B 78 (2008) 165420.

[19] S. E. Savelev, W. Hausler and P. Hanggi, Phys. Rev. Lett. 109 (2012) 226602.

[20] W. T. Lu, et. al. J. Appl. Phys. 111 (2012) 103717.

[21] H. Y. Chen, Phys. Let. A 378(2014) 2226

[22] R. Zhu, J. H. Dai and Y. Guo, J. Appl. Phys. 117 (2015) 164306.

[23] A. R. Wright, X. G. Xu, J. C. Cao and C. Zhang, Appl. Phys. Lett. 95 (2009) 072101.

[24] R. Biswas, S. Maity and C. Sinha, Physica E: 84 (2016) 235.

[25] Q. S. Wu, S. N. Zhang and S. J. Yang, J. Phys.: Condens. Matter 20 (2008) 485210.

[26] N. Myoung and G. Ihm, Physica E 42 (2009) 70.

[27] M. R. Masir, P. Vasilopoulos and F. M. Peeters, New J. Phys. 11 (2009) 095009.

[28] A. V. Shytov, M. S. Rudner and L. S. Levitov, Phys. Rev. Lett. 101 (2008) 156804.

[29] M. R. Masir, P. Vasilopoulos and F. M. Peeters, Phys. Rev. B 82 (2010) 115417.

[30] J. H. Shirley, Phys. Rev. 138 (1965) B979.

[31] W. Li and L. E. Reichl, Phys. Rev. B 60 (1999) 15732.

[32] Z. Gu, H. A. Fertig, D. P. Arovas and A. Auerbach, Phys. Rev. Lett. 107 (2011) 216601.

[33] B. Trauzettel, Ya. M. Blanter and A. F. Morpurgo, Phys. Rev. B 75 (2007) 035305.

[34] M. V. Fistul and K. B. Efetov, Phys. Rev. Lett. 98 (2007) 256803

[35] A. Iurov, G. Gumbs, O. Roslyak and D. Huang, J. Phys.: Condens. Matter 24 (2012) 015303.

[36] H. L. Calvo, H. M. Pastawski, S. Roche and L. E. F. F. Torres, Appl. Phys. Lett. 98 (2011) 232103; ibid, 101 (2012) 253506.

[37] C. Sinha and R. Biswas, Appl. Phys. Lett. 100 (2012) 183107.

[38] R. Biswas and C. Sinha, J. Appl. Phys. **114** (2013) 183706.

[39] U. Fano, Phys. Rev. **124** (1961) 1866.

[40] L. Szabo, M. G. Benedict, A. Czirjak and P. Foldi, Phys. Rev. B 88 (2013) 075438.

[41] C. Zhang, J. Liu and L. Fu, Euro. Phys. Lett. 110 (2015) 61001.

[42] R. Biswas and C. Sinha, J. Appl. Phys., **115** (2014) 133705.

[43] M. V. Fistul and K. B. Efetov, Phys. Rev. B 90 (2014) 125416.

[44] B. Lukyanchuk et. al., Nat. Mater. 9 (2010) 707.

[45] I. S. Gradshteyn and I. M. Ryzhik, Table of Integrals, Series, and Product (Academic Press, Inc. NY, 1980).

Figure Captions:

Fig.1: (a) Vector potential profile corresponding to a rectangular magnetic barrier of width 'L' and height 'B'. (b) The magnetic field profile in the three regions (I, II and III) corresponding to the potential of Fig.(a). (c) Sinusoidally varying time dependent scalar potential of amplitude $V_0$ and frequency $\omega$ applied in region II. (d) Magnetic field profile corresponding to a pair of $\delta$- function magnetic barriers of strength '$B$' and separated by a distance '$L$'. (e) Magnetic vector potential profile corresponding to the inhomogeneous magnetic field shown in (d).

Fig.2: Transmission coefficient $T_c$ plotted as a function of incident energy ($E$) for '$k_y$'= 0.5 and '$L$' = 20. (a) For static $\delta$-magnetic barriers with '$B$' = 2 and (b) for static electric barrier with barrier height '$V_z$' = 2.

Fig.3: (Color online only) Total side band transmission $T_c$ ($\sum_m T_m$) plotted as a function of incident energy ($E$) for '$B$'=2, $k_y$= 0.5 and '$L$' = 20.0, '$V_0$' = 1 and '$\omega$' = 0.5. Dash (black) line for static (FF) vector barrier and solid (red) line for oscillating vector barrier.
Inset: Oscillating scalar barrier (SB) with barrier height '$V_z$' = 2.0.

Fig.4: (Color online only) Same as Fig. 3 but '$B$'=2, '$L$' = 20.0, $k_y$= 0.5, '$V_0$' = 0.5. (a) dash (black) for static vector barrier (FF) and solid (red) for '$\omega$' = 0.1, (b) for '$\omega$' = 0.2, (c) for '$\omega$' = 0.3, and (d) for '$\omega$' = 0.4.

Fig.5: (Color online only) Same as Fig. 3 but '$L$' = 20.0, $k_y$= 0.5, '$V_0$' = 1 and '$\omega$' = 0.6; Inset: for , '$k_y$'= - 0.5.

Fig.6: (Color online only) Same as Fig. 3 but '$B$' = 1, '$L$' = 2.0 and '$k_y$' = -1. (a) for '$V_0$' = 0.5 and '$\omega$' = 0.8 for solid (black), 0.7 for dash (red), and 0.6 for dot (blue). (b) for '$\omega$' = 0.7

and '$V_0$' = 1.0 for solid (black), 0.8 for dash (red), 0.5 for dot (blue), and 0.0 for dash-dot (dark-cayn).

Fig.7: (Color online only) Same as Fig.6 but $k_y$= - 0.5. (a) for '$V_0$' = 0.8 and '$\omega$' = 0.6 for solid (black), 0.65 for dash (red), and 0.7 for dot (blue). (b) for '$\omega$' = 0.71 and '$V_0$' = 0.6 for solid (black), 0.8 for dash (red), and 1.0 for dot (blue). Inset: zoom around '$E$' = 0.92.

Fig.8: (Color online only) Transmission coefficient ($T_c$) for individual Floquet band (e.g., no photon ($T_0$), single photon absorption ($T_{+1}$), two photon absorption ($T_{+2}$), single photon emission ($T_{-1}$), and two photon emission ($T_{-2}$)) as a function of energy. (a) '$B$' = 2, '$L$' = 20.0, $k_y$= 0.2, '$V_0$' = 1.0, and '$\omega$' = 0.6. Solid (black) line for $T_0$, dash (red) line for $T_{+1}$, and dot (blue) for $T_{-1}$. Inset: Solid (black) for $T_{+2}$ and dot (red) for $T_{-2}$. (b) Same as (a) but for '$V_0$' = 0.5 and '$\omega$' = 0.6. (c) '$B$' = 1, '$L$' = 2.0, $k_y$= -1, '$V_0$' = 0.4, and '$\omega$' = 0.7. Solid (black) line for $T_0$, dash (red) line for $T_{+1}$ and dot (blue) line for $T_{-1}$.

Fig.9: (Color online only) Total side band transmission $T_c$ ($\sum_m T_m$) plotted as a function of (a) incident angle for '$B$' = 1.0, '$E$'= 1.0, '$\omega$' = 0.9 and '$L$' = 1.0 and (b) incident energy for $k_y$= 0.0, '$L$' = 1.5, '$V_0$' = 1 and '$B$' = 1.0. (a) Solid (black) line for static (FF) vector barrier, Dash (red) line for '$V_0$' = 0.2, dot (blue) for '$V_0$' = 0.5 and dash-dot (green) line for '$V_0$' = 0.7. (b) Solid(black) line for FF, dash (red) line for '$\omega$' = 0.4, dot (blue) line for '$\omega$' = 0.5 and dash-dot (black) line for '$\omega$' = 0.6.

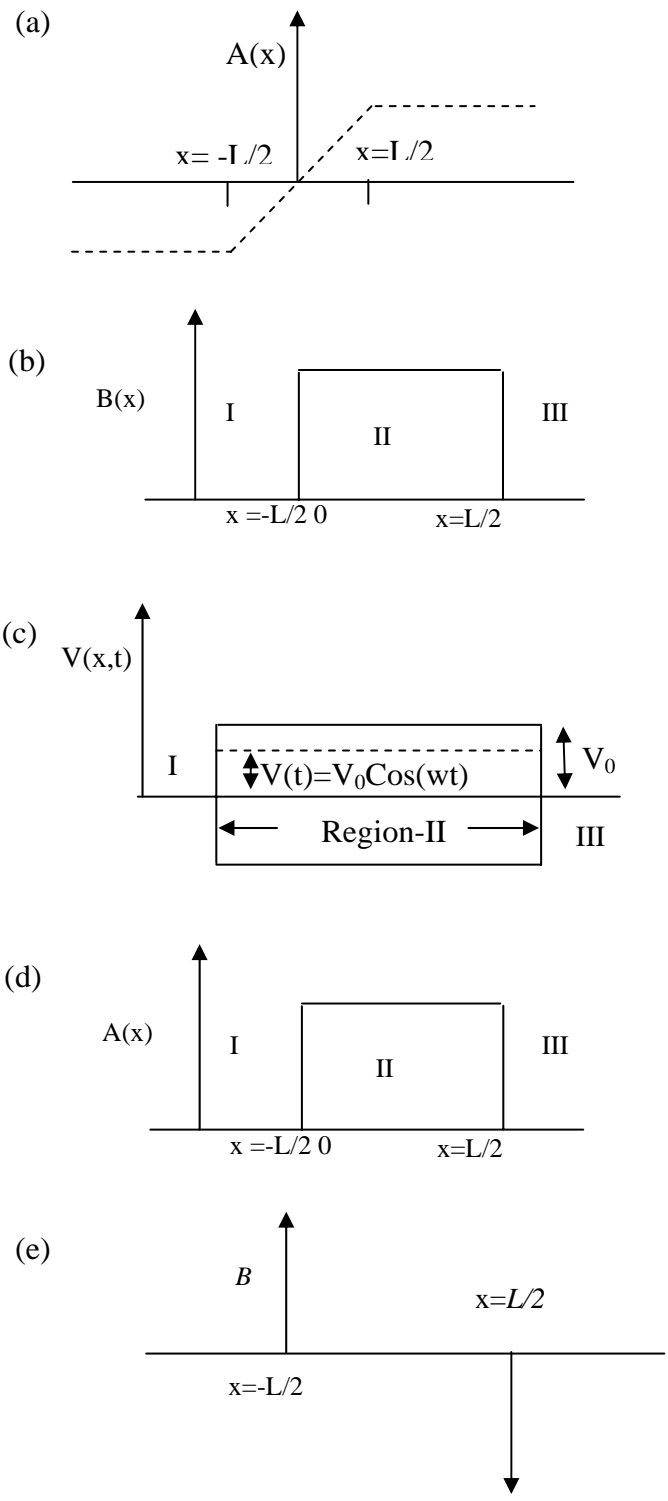

Fig. 1

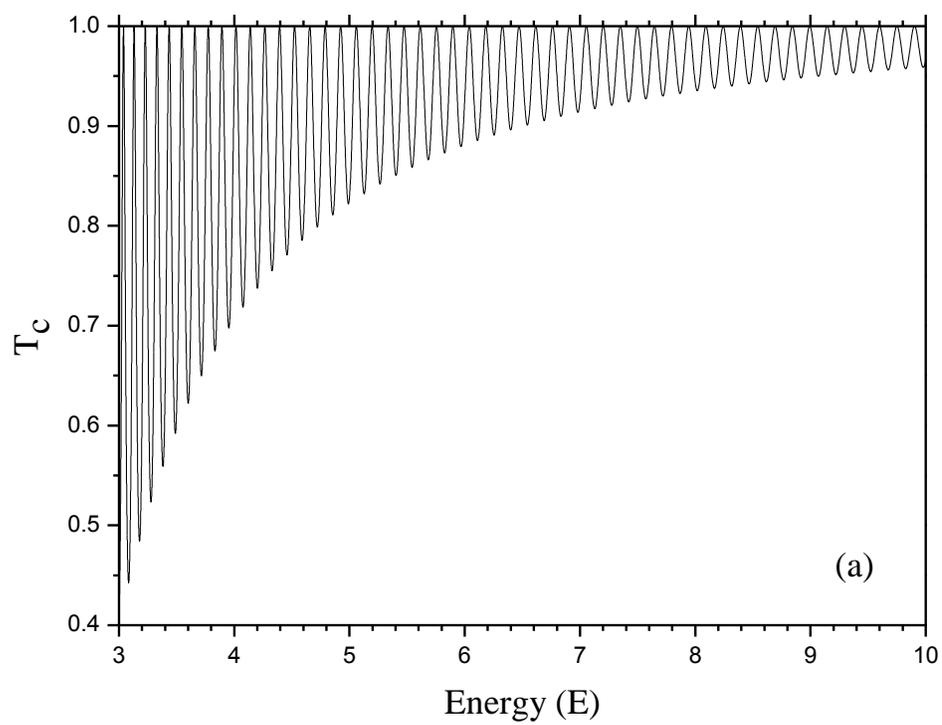

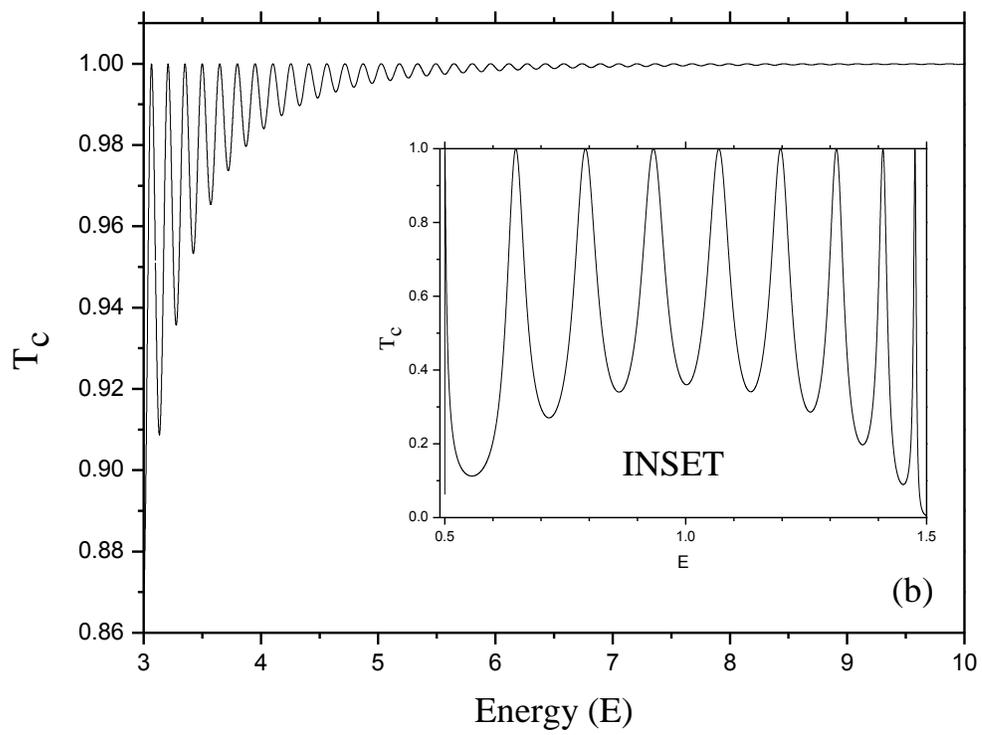

Fig.2

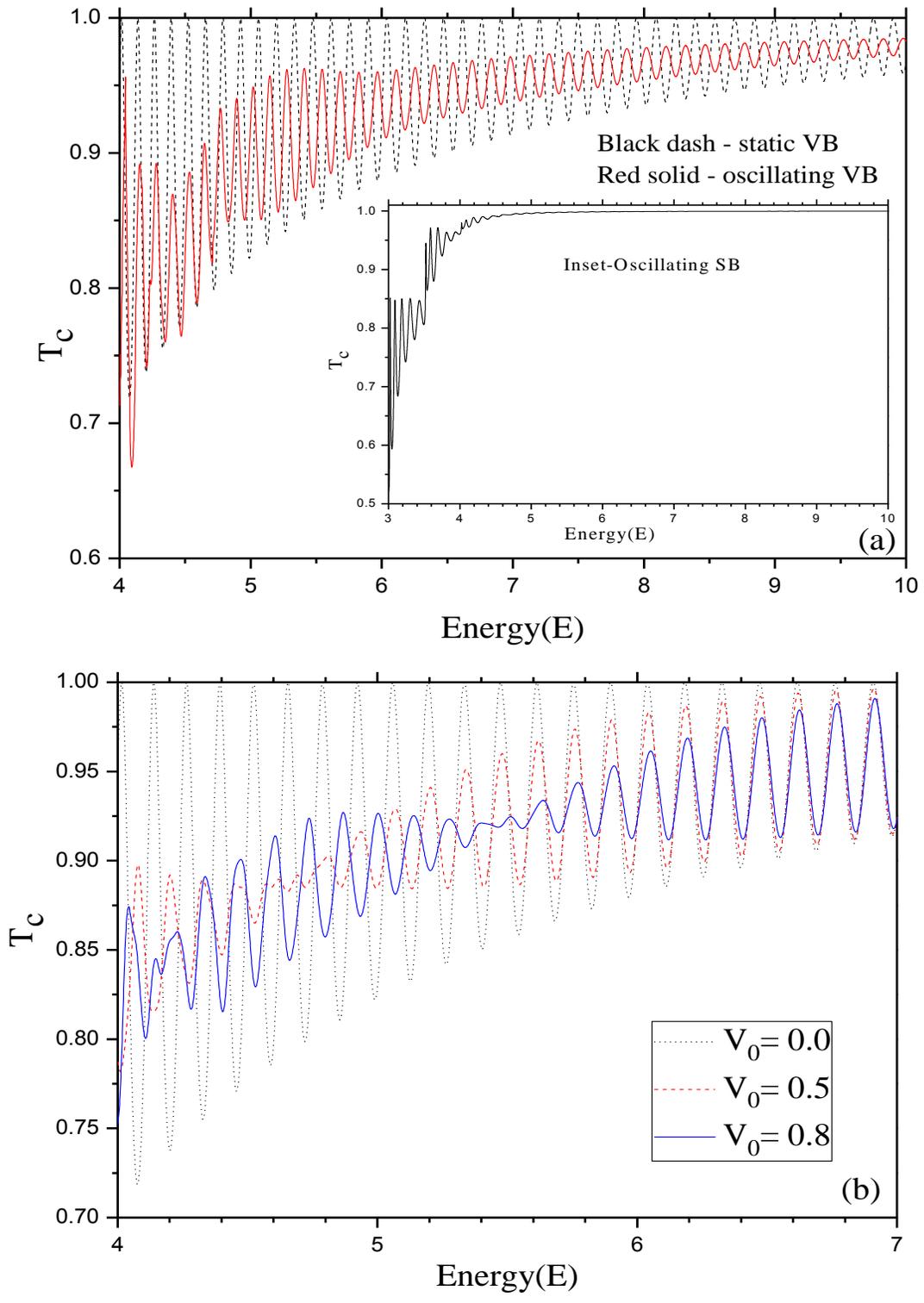

Fig.3

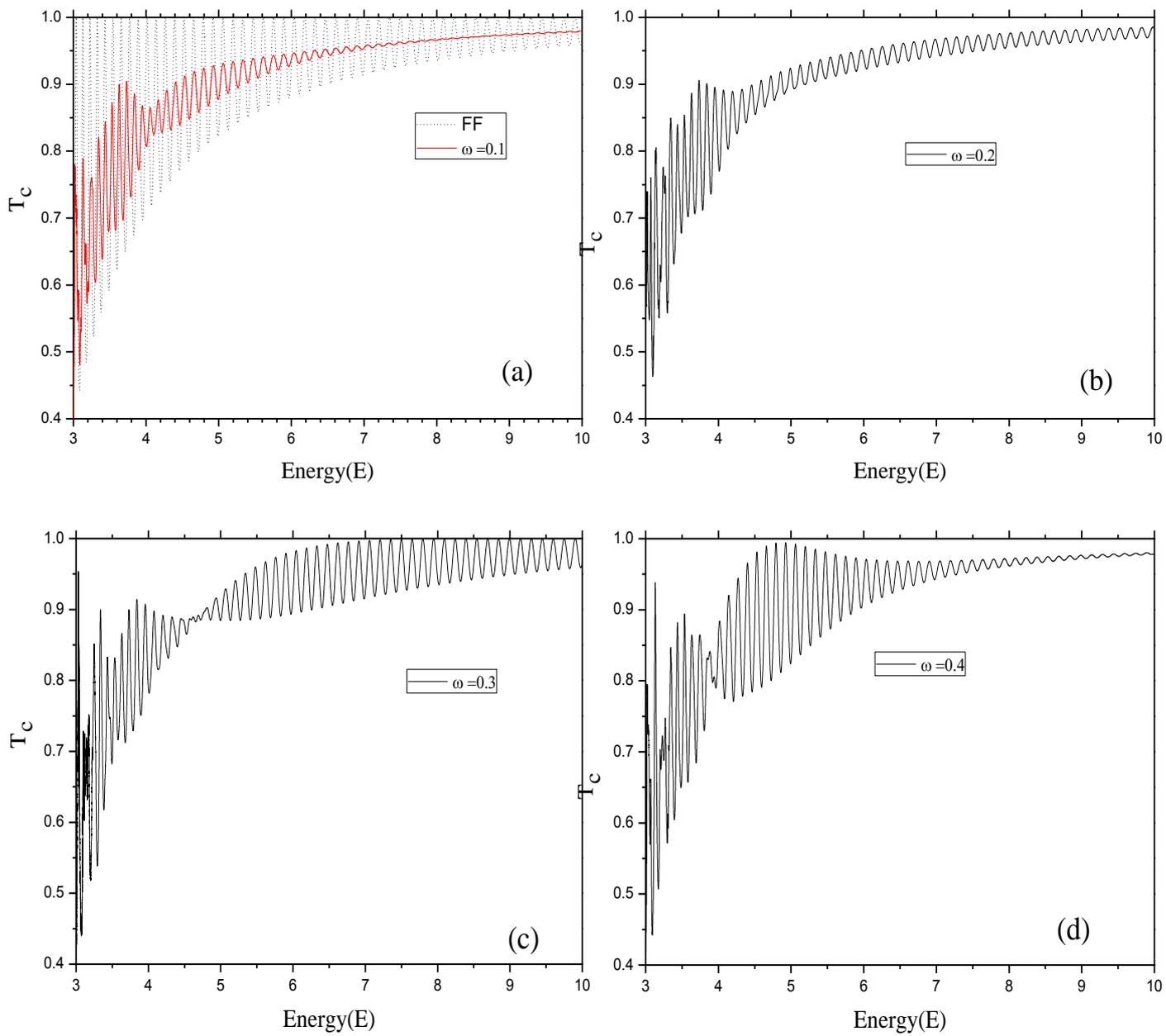

Fig.4

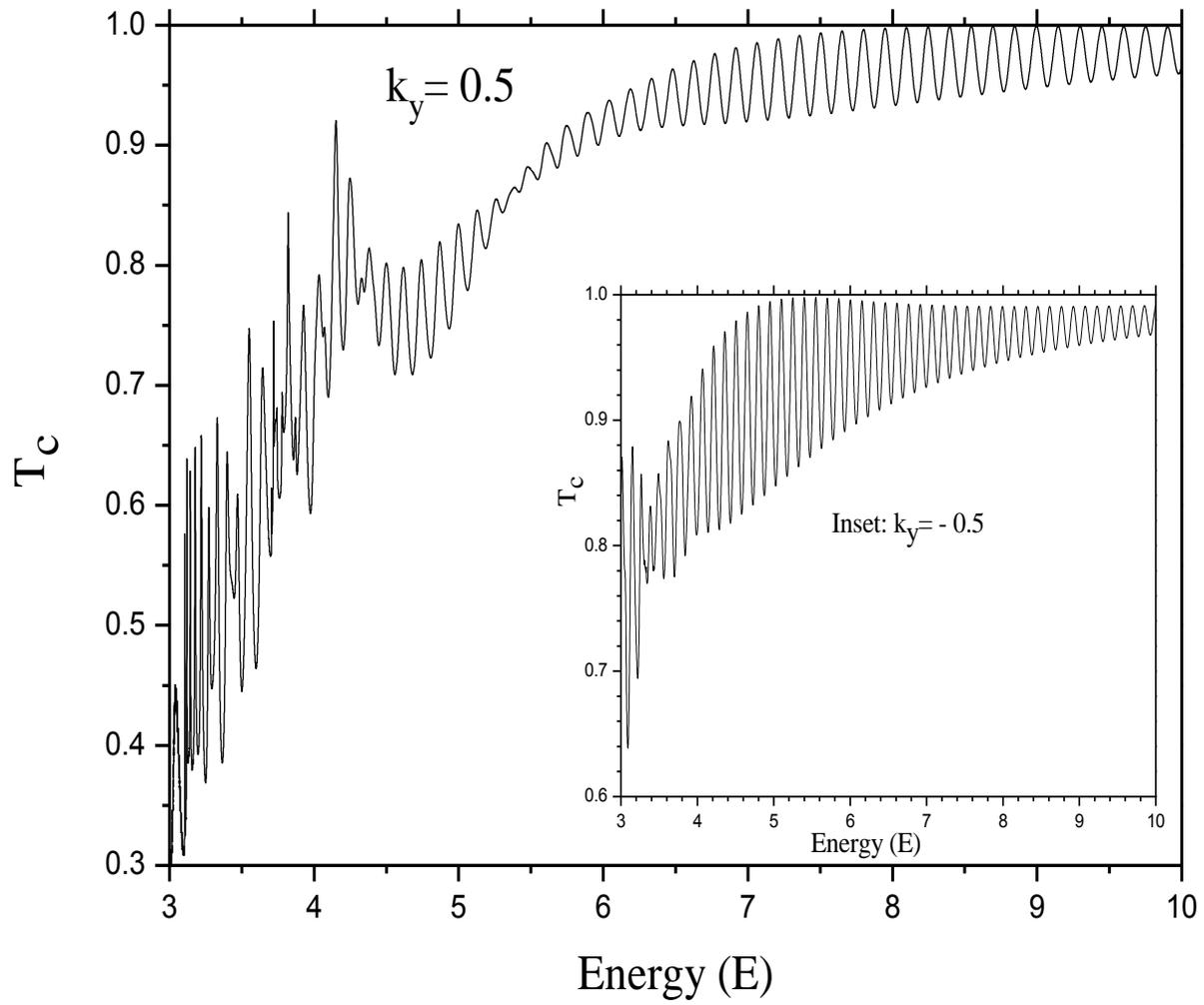

Fig.5

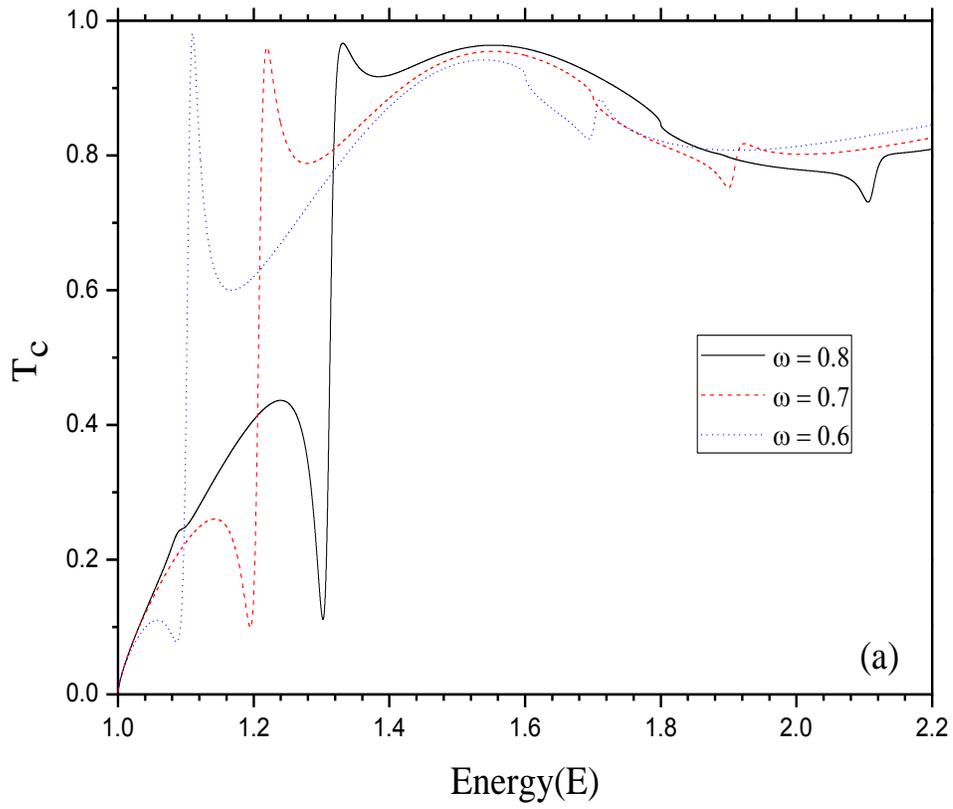

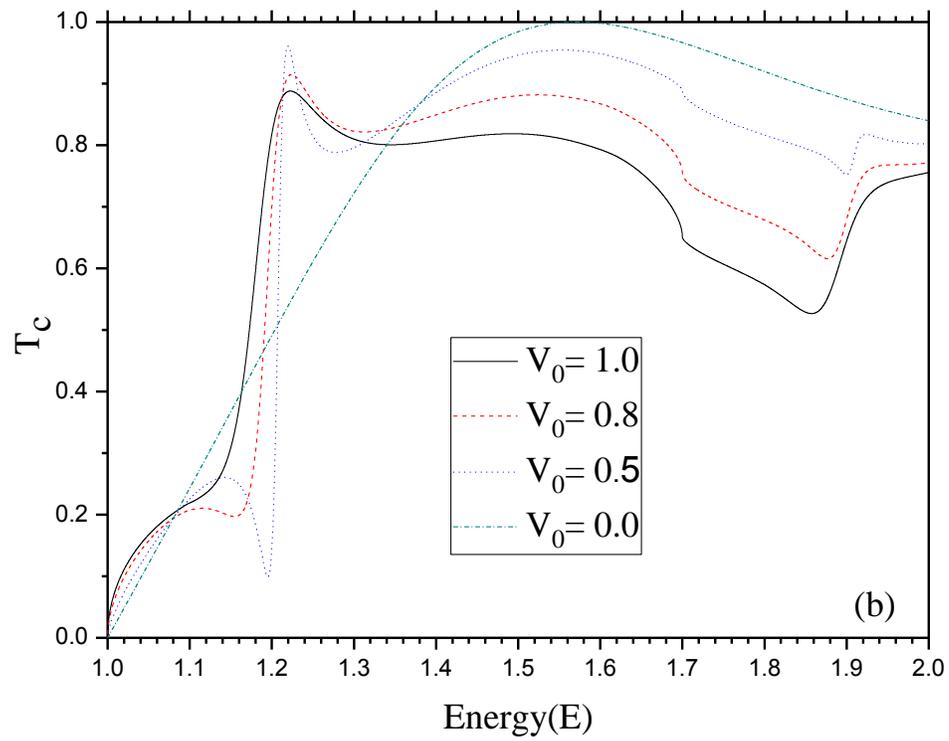

Fig.6

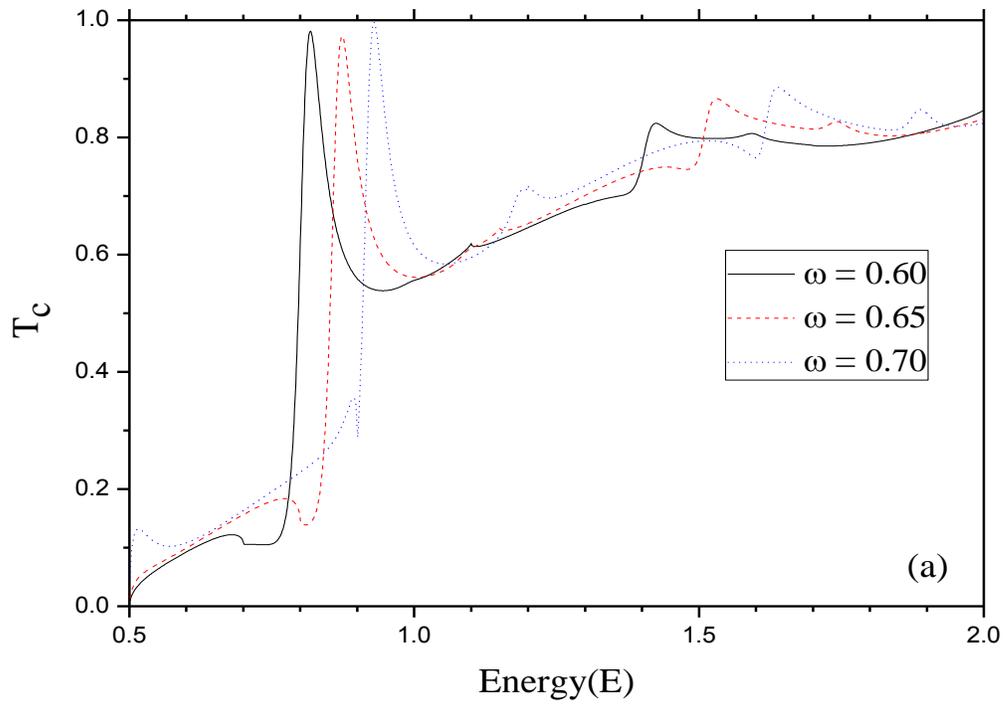

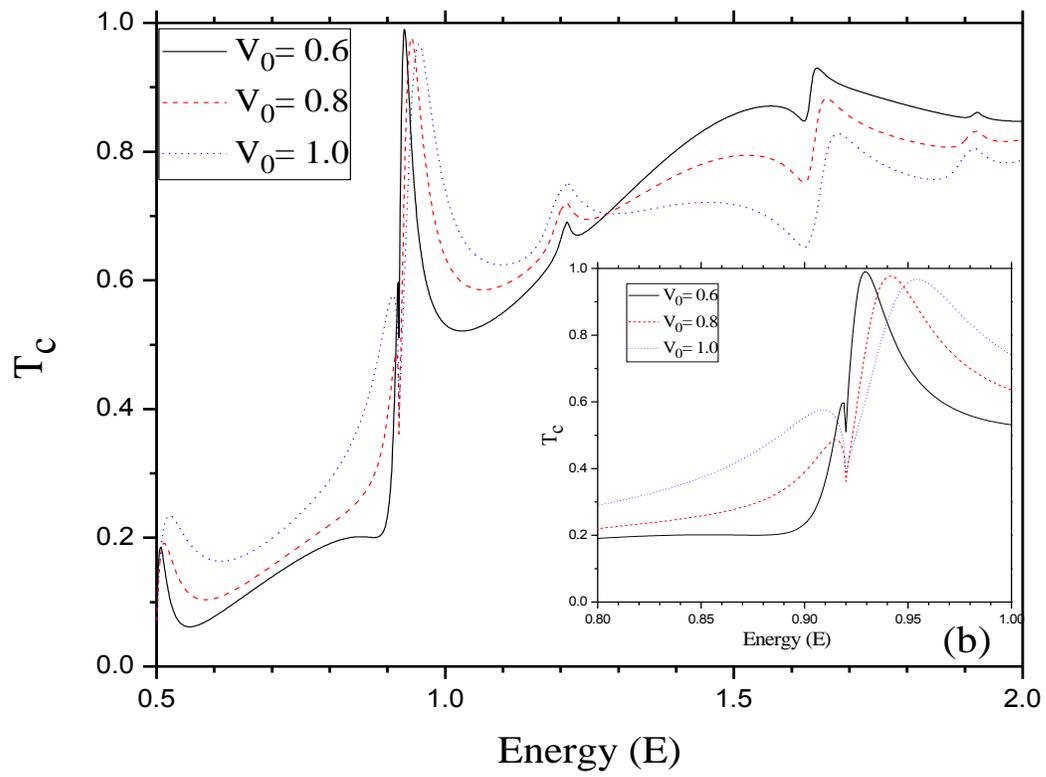

Fig.7

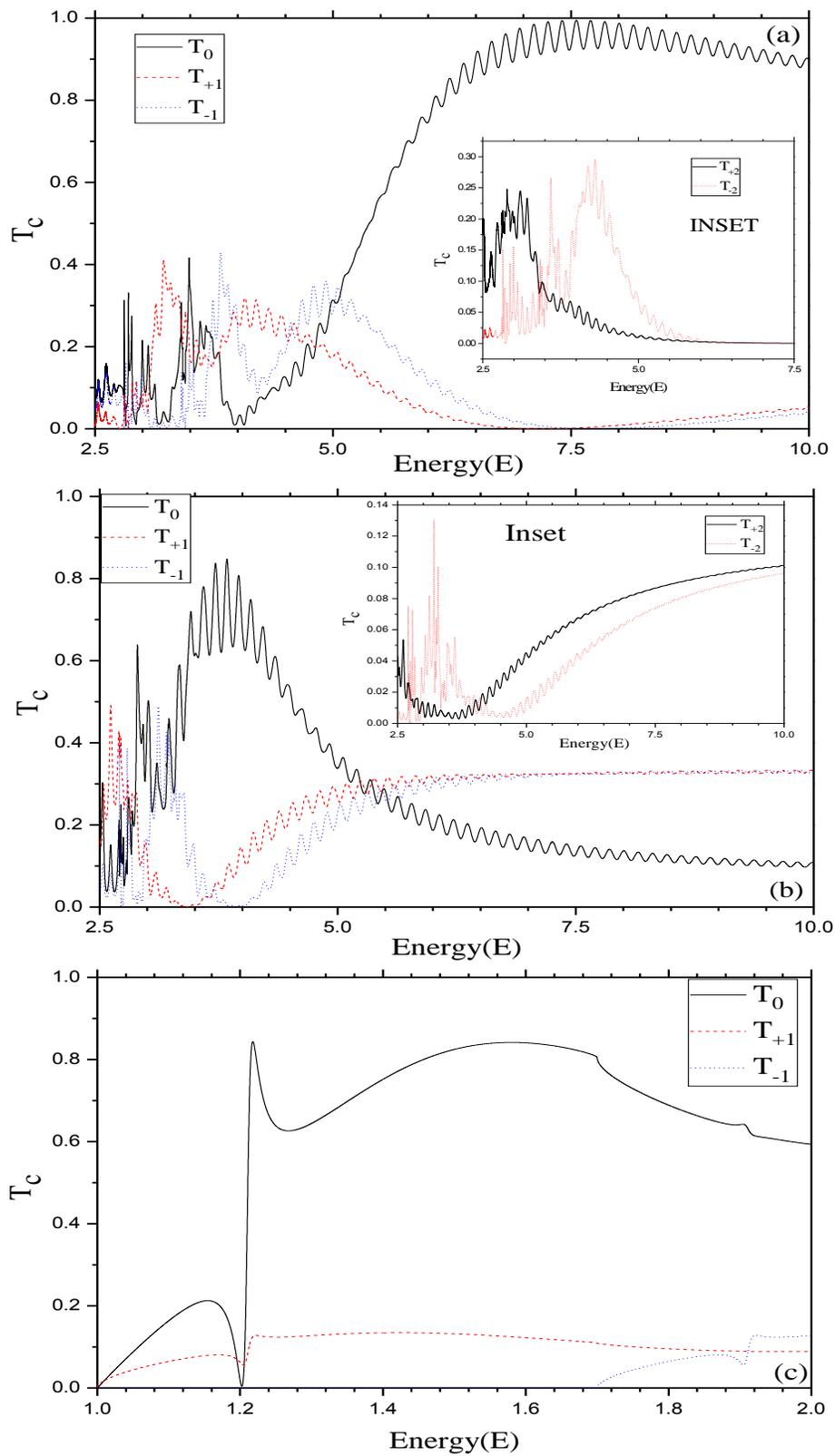

Fig.8

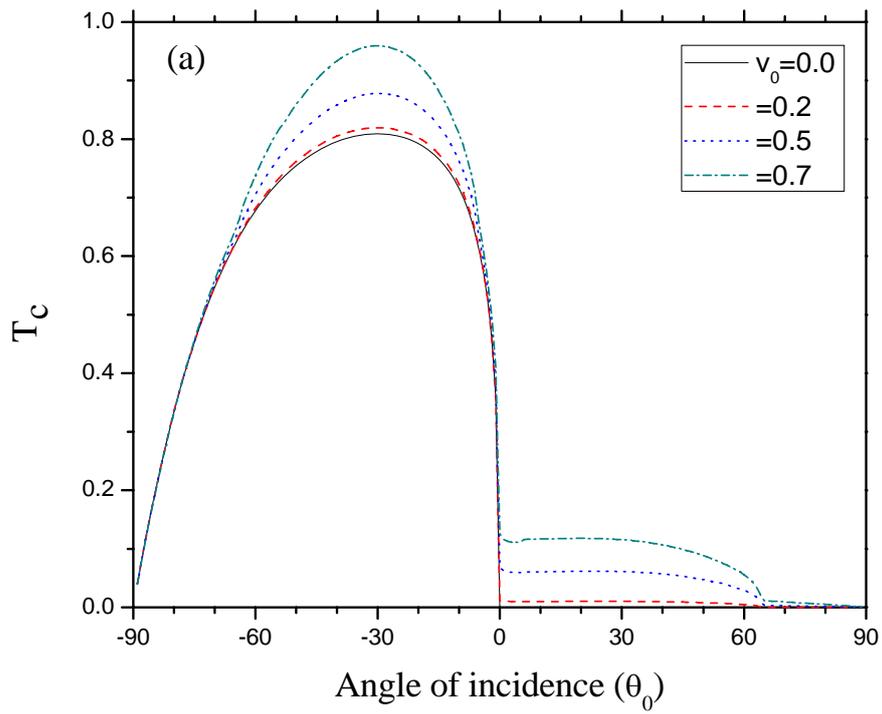

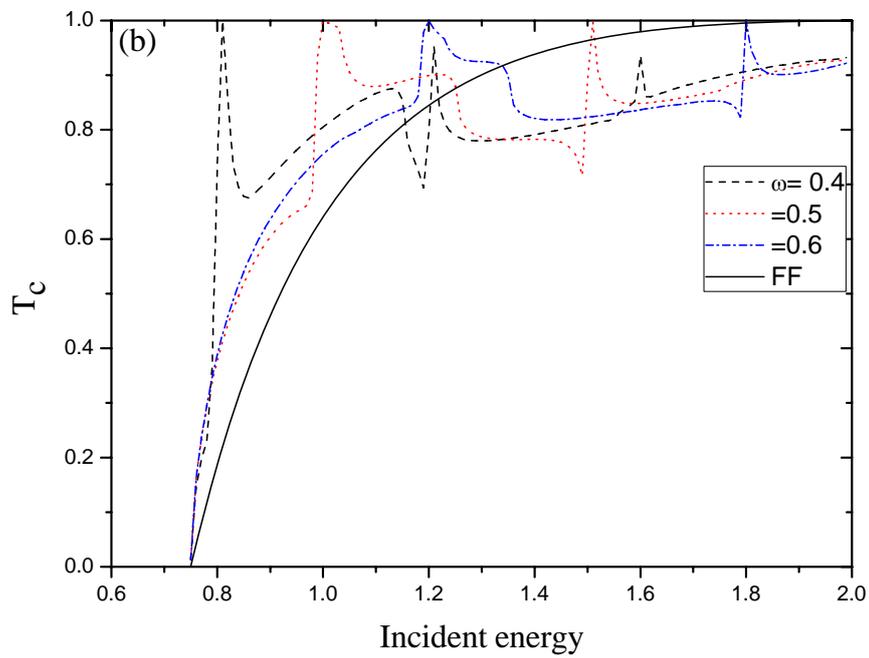

Fig. 9